\documentclass[prc,twocolumn,showpacs,superscriptaddress]{revtex4}
\usepackage{rotating}
\usepackage{amsmath,amssymb}
\usepackage{bm,subfigure}
\usepackage{verbatim}
\usepackage{graphicx}
\usepackage{dcolumn}
\usepackage{amstext}
\usepackage{lscape}
\usepackage{mathptmx}
\usepackage{verbatim}
\usepackage[colorlinks,citecolor=blue,bookmarks]{hyperref}
\usepackage{times}

\begin{document}

\title{Examining the stability of thermally fissile Th and U isotopes}
 
\author{Bharat Kumar, S. K. Biswal, S. K. Singh and S. K. Patra}

\affiliation{
Institute of Physics, Bhubaneswar-751005, India}
\date{\today}

\begin{abstract}
The properties of recently predicted thermally fissile Th and U isotopes
are studied within the framework of relativistic mean field (RMF) 
approach using  axially deformed basis. We calculated the ground,  
first intrinsic excited state for highly neutron-rich 
thorium and uranium isotopes. The possible modes of decay like 
$\alpha$-decay and $\beta$-decay are analyzed. We found that the 
neutron-rich isotopes are stable against 
$\alpha$-decay, however they are very much unstable against $\beta$-decay.
The life time of these nuclei predicted to be tens of second against 
$\beta$-decay. If these nuclei utilize before their decay time, a lots of 
energy can be produced with the help of multi-fragmentation fission. Also,
these nuclei have a great implication in astrophysical point of view.  
In some cases, we found the isomeric states with energy range
from 2 to 3 MeV and three maxima in the potential energy surface of 
$^{228-230}$Th and $^{228-234}$U isotopes.
\end{abstract}
\maketitle
\footnotetext [1]{bharat@iopb.res.in}
\section{Introduction}\label{sec1}

Now-a-days uranium and thorium isotopes have attracted a great attention 
in nuclear physics due to the thermally fissile nature of some of its  
isotopes\cite{sat08}. These thermally fissile materials have tremendous 
importance in energy production. Till date, the known thermally fissile nuclei 
are $^{233}$U, $^{235}$U and $^{239}$Pu. Out of which only $^{235}$U has a 
long life time and the only thermally fissile isotope available in nature  
\cite{sat08}. Thus, presently it is an important area of research
to look for any other thermally fissile nuclei apart from $^{233}$U,
 $^{235}$U and $^{239}$Pu. Recently, Satpathy et al. 
\cite{sat08} showed that uranium and thorium  isotopes with neutron number 
N=154-172 have thermally fissile property. They performed 
a calculation with a typical example of $^{250}$U that this nucleus has a 
low fission barrier with a significantly large barrier width, which makes it
stable against the spontaneous fission. Apart from the thermally fissile 
nature, these nuclei also play an important role in the nucleosynthesis 
in the stellar evolution. As these nuclei are stable against 
spontaneous fission, thus the prominent decay
modes may be the emission of $\alpha$-, $\beta$- and $cluster$-particles 
from the neutron-rich thermally fissile (uranium and thorium) isotopes.

To measure the stability of these neutron-rich U and Th isotopes, 
 we investigated the $\alpha$- and $\beta$- decay properties
of these nuclei. Also, we extend our calculations to estimate the 
binding energy, root mean square  radii, quadrupole moments and 
other structural properties. 

From last three decades, the  relativistic mean field (RMF) formalism
is a formidable theory in describing the finite nuclear properties throughout
 the periodic chart and infinite nuclear matter properties concerned 
with the cosmic dense object like neutron star. In the same line RMF theory 
is also good enough to study the clusterization \cite{aru05}, $\alpha$-decay 
\cite{bidhu11}, and $\beta$-decay of nuclei. The presence of cluster 
in  heavy nuclei like, $^{222}$Ra, $^{232}$U, $^{239}$Pu and 
$^{242}$Cm has been studied using RMF formalism \cite{bk06,patra07}. 
It gives a clear prediction of $\alpha$-like (N=Z) matter at the central part 
for heavy nuclei and $cluster$-like structure (N=Z and $N\neq Z$) for light 
mass nuclei \cite{aru05}. The proton emission as well as the cluster decay 
phenomena are well studied using RMF formalism  with M3Y~\cite{love79},  
LR3Y~\cite{bir12} and NLR3Y\cite{bidhu14} nucleon-nucleon potentials in 
the framework of single and double folding models, respectively. 
Here, we used the relativistic mean field (RMF) formalism 
with the well known NL3 parameter set \cite{lala97}  for all our calculations.

The paper is organized as follows: The RMF formalism is outlined briefly
in Section~\ref{sec2}. The importance of pairing correlation and  inclusion with
BCS approximation  are also given in this section. 
The results obtained from our calculations for binding energy, basis selection,
potential energy surface (PES) diagrams and the evaluation of single-particle
levels are discussed in Section~\ref{sec4}. 
The $Q_{\alpha}$- and $Q_{\beta}$-values are calculated in section~\ref{sec9}.
 In this section, various decay modes  are discussed using  
empirical formula and limitation of the model is also given same section.  
Finally, a brief summary and concluding remarks are given in the last
 Section~\ref{sec14}.

\section{Formalism}\label{sec2}
In present manuscript, we used the axially deformed relativistic 
mean field formalism to calculate various nuclear phenomena. The 
meson-nucleon interaction is given by  ~\cite{patra91,wal74,seort86,horo 81,
bogu77,pric87} 
\begin{eqnarray}
{\cal L}&=&\overline{\psi_{i}}\{i\gamma^{\mu}
\partial_{\mu}-M\}\psi_{i}
+{\frac12}\partial^{\mu}\sigma\partial_{\mu}\sigma
-{\frac12}m_{\sigma}^{2}\sigma^{2}\nonumber\\
&& -{\frac13}g_{2}\sigma^{3} -{\frac14}g_{3}\sigma^{4}
-g_{s}\overline{\psi_{i}}\psi_{i}\sigma-{\frac14}\Omega^{\mu\nu}
\Omega_{\mu\nu}\nonumber\\
&&+{\frac12}m_{w}^{2}V^{\mu}V_{\mu}
+{\frac14}c_{3}(V_{\mu}V^{\mu})^{2} -g_{w}\overline\psi_{i}
\gamma^{\mu}\psi_{i}
V_{\mu}\nonumber\\
&&-{\frac14}\vec{B}^{\mu\nu}.\vec{B}_{\mu\nu}+{\frac12}m_{\rho}^{2}{\vec
R^{\mu}} .{\vec{R}_{\mu}}
-g_{\rho}\overline\psi_{i}\gamma^{\mu}\vec{\tau}\psi_{i}.\vec
{R^{\mu}}\nonumber\\
&&-{\frac14}F^{\mu\nu}F_{\mu\nu}-e\overline\psi_{i}
\gamma^{\mu}\frac{\left(1-\tau_{3i}\right)}{2}\psi_{i}A_{\mu} .
\label{eq:7}
\end{eqnarray}
Where, $\psi$ is the Dirac spinor and meson fields are denoted by 
$\sigma$,$V^\mu$ and $R^\mu$ for $\sigma$,$\omega$ and $\rho-$ meson  
respectively. The electromagnetic interaction between the proton is denoted 
by photon field $A^\mu$. $g_s$, $g_\omega$, $g_\rho$ and $\frac{e^2}{4\pi}$ 
are the coupling constants for the $\sigma$,$\omega$ and $\rho-$ meson 
and photon field respectively. The strength of the self coupling $\sigma-$ 
meson ($\sigma^3$ and $\sigma^4$) are denoted by $g_2$ and $g_3$, along with 
$c_3$ as the non-linear coupling constant for $\omega$ meson. 
The nucleon mass is scripted as M, where the $\sigma$, $\omega$, and 
$\rho-$ meson masses are $m_s$, $m_\omega$ and $m_\rho$ respectively.
From the classical Euler-Lagrangian equation, we get the Dirac-equation 
and Klein- Gordan equation for the nucleon and meson field respectively. 
The Dirac-equation for the nucleon is solved by expanding the Dirac 
spinor into lower and upper component, while the mean field equation 
for the Bosons are solved in deformed harmonic oscillator basis 
with $\beta_0$ as the deformation parameter. The nucleon equation 
along with different meson equation form a coupled set of equation, 
which  can be solved by iterative method. Various types of densities 
such as baryon (vector), scalar, isovector and proton (charge) densities 
are given as
\begin{eqnarray}
\rho(r) & = &
\sum_i \psi_i^\dagger(r) \psi_i(r) \,,
\label{eqFN6} \\[3mm]
\rho_s(r) & = &
\sum_i \psi_i^\dagger(r) \gamma_0 \psi_i(r) \,,
\label{eqFN7} \\[3mm]
\rho_3 (r) & = &
\sum_i \psi_i^\dagger(r) \tau_3 \psi_i(r) \,,
\label{eqFN8} \\[3mm]
\rho_{\rm p}(r) & = &
\sum_i \psi_i^\dagger(r) \left (\frac{1 -\tau_3}{2}
\right)  \psi_i(r) \,.
\label{eqFN9} % \\[3mm]
\end{eqnarray}
The calculations are simplified under the shadow of various symmetries like 
conservation of parity, no-sea approximation  and time reversal symmetry, 
which kills all spatial components of the meson fields and the anti-particle 
states contribution to nuclear observable. The center of mass correction is 
calculated with the non-relativistic approximation, which gives 
$E_{c.m}=\frac{3}{4}  41A^{-1/3}$ (in MeV).
The quadrupole deformation parameter $\beta_2$ is calculated from the resulting 
quadropole moments of the proton and neutron. The 
binding energy and charge radius are given by well known 
relation \cite{blunden87,reinhard89,gam90}. 

\subsection{Pairing correlations in RMF formalism}\label{sec3}

In  nuclear structure physics, the pairing correlation has  an 
indispensable role in open shell nuclei. The priority of the pairing 
correlation escalates with mass number A. It also plays a crucial role for the 
understanding of deformation of heavy nuclei. Because of the limited
pair near the Fermi surface, it has a nominal effect for light mass nuclei
on both bulk and single-particle properties.  In the present case, we consider
only T=1 channel of pairing correlation, i.e., pairing between proton-proton 
and neutron-neutron. In such case, a nucleon of quantum state $|j, m_z\rangle$
pairs with another nucleon having same $I_z$ value with quantum state 
$|j,{-m_z}\rangle$, which is the time reversal partner of other. The 
philosophy of BCS pairing is same both in nuclear and atomic domain. The 
first evidence of the pairing energy came from the even-odd mass staggering 
of isotopes. In mean field formalism  the violation of particle number is 
account of pairing correlation.
The RMF Lagrangian density only accommodates term like 
${\psi}^{\dagger}{\psi}$ (density) and no term of the form 
${\psi}^{\dagger}{\psi}^{\dagger}$ or $\psi\psi$. The inclusion 
of pairing correlation of the form $\psi \psi$ or 
${\psi}^{\dagger}{\psi}^{\dagger}$ violates the particle number 
conservation~\cite{patra93}. Thus, a constant gap BCS-type simple prescription 
is adopted in our calculations to take care of the pairing correlation
for open shell nuclei.  The general expression 
for pairing interaction to the total energy in terms of occupation 
probabilities $v_i^2$ and $u_i^2=1-v_i^2$ is written 
as~\cite{pres82,patra93}:
\begin{equation}
E_{pair}=-G\left[\sum_{i>0}u_{i}v_{i}\right]^2,
\end{equation}
with $G=$ pairing force constant. 
The variational approach with respect to the occupation number $v_i^2$ 
gives the BCS equation 
\cite{pres82}:
\begin{equation}
2\epsilon_iu_iv_i-\triangle(u_i^2-v_i^2)=0,
\label{eqn:bcs}
\end{equation}
with $\triangle=G\sum_{i>0}u_{i}v_{i}$. 

The densities with occupation number is defined as:
\begin{equation}
n_i=v_i^2=\frac{1}{2}\left[1-\frac{\epsilon_i-\lambda}{\sqrt{(\epsilon_i
-\lambda)^2+\triangle^2}}\right].
\end{equation}
For the pairing gap ($\triangle$) of proton and neutron is taken from 
the phenomenological formula of Madland and Nix \cite{madland}:
\begin{eqnarray}
\triangle_n=\frac{r}{N^{1/3}}exp(-sI-tI^{2})
\\
\triangle_p=\frac{r}{Z^{1/3}}exp(sI-tI^{2})
\end{eqnarray}
where, $I=(N-Z)/A$, $r=5.73$ MeV, $s=0.117$, and $t=7.96$.
 
The chemical potentials $\lambda_n$ and $\lambda_p$ are determined by the
particle numbers for neutrons and protons. The pairing energy of the 
nucleons using  equation (7) and (8) can be written as:
\begin{equation}
E_{pair}=-\triangle\sum_{i>0}u_{i}v_{i}.
\end{equation}

In constant pairing gap calculation, for a particular value of 
pairing gap $\triangle$ and force constant $G$, the pairing energy $E_{pair}$
diverges, if it is extended to an infinite configuration space.
In fact, in all realistic calculations with finite range forces,
the contribution of states of large momenta above the Fermi surface
(for a particular nucleus) to  $\triangle$ decreases with energy.
Therefore, the pairing window in all the equations are extended upto the level 
$|\epsilon_i-\lambda|\leq 2(41A^{-1/3})$ as a function of the single 
particle energy. 
The factor 2 has been determined so as to reproduce the pairing correlation 
energy for neutrons in $^{118}$Sn using Gogny force 
\cite{gam90,patra93,dech80}.  We notice that recently Karatzikos 
et al.~\cite{karatzikos10} has been shown that if it is adjusted a
constant pairing window for a particular deformation then it may leads to 
errors at different energy solution (different state solution). 
However, this kind of approach have not taken into account in our calculations, 
as we have adjusted to reproduce the pairing as a whole for $^{118}Sn$ nucleus.

It is a tough task to compute the binding energy and
quadrupole moment of odd-N or odd-Z or both N and Z numbers
are odd (odd-even, even-odd, or odd-odd) nuclei. 
To do this, one needs to include the additional
time-odd term, as is done in the SHF Hamiltonian~\cite{stone07}, or 
empirically the pairing force in order to take care the effect 
of odd-neutron  or odd-proton~\cite{onsi20}. In an odd-even or odd-odd nucleus,
 the time reversal symmetry gets violated in the mean field models.
In our RMF calculations, we neglect the space components
of the vector fields, which are odd under time reversal and
parity. These are important in the determination of magnetic
moments~\cite{holf88} but have a very small effects on bulk properties
such as binding energies or quadrupole deformations, and they
can be neglected\cite{lala99} in the present context. Here, for the
odd-Z or odd-N calculations, we employ the Pauli blocking approximation,
which restores the time-reversal symmetry. In this approach,
one pair of conjugate states, $\pm{m}$, is taken out of the pairing
scheme. The odd particle stays in one of these states, and its
corresponding conjugate state remains empty. In principle, one
has to block in turn different states around the Fermi level to
find the one that gives the lowest energy configuration of the
odd nucleus. For odd-odd nuclei, one needs to block both the
odd neutron and odd proton.

\section{Calculations and results}\label{sec4}

In this Section, we evaluate our results for binding energy, rms radii,
quadrupole deformation parameter for recently predicted thermally fissile 
isotopes of  Th and U.  These nuclei are quite heavy and needed a large
number of oscillator basis, which takes considerable time for computation.
We spent few lines in the first subsection of this section to 
describe how to select basis space and the results and discussions are followed 
subsequently.

\subsection{Selection of basis space}\label{sec5}

The Dirac equation for Fermions (proton and neutron) and the equation of 
motion for Bosons ($\sigma-$, $\omega-$, $\rho-$ and $A_0$) obtained from
the RMF Lagrangian are solved self-consistently using an iterative methods.
These equations are solved in an axially  deformed harmonic oscillator 
expansion basis $N_F$ and $N_B$ for Fermionic and Bosonic wavefunction,
respectively.

\begin{figure}[ht]
\includegraphics*[width=1.1\columnwidth]{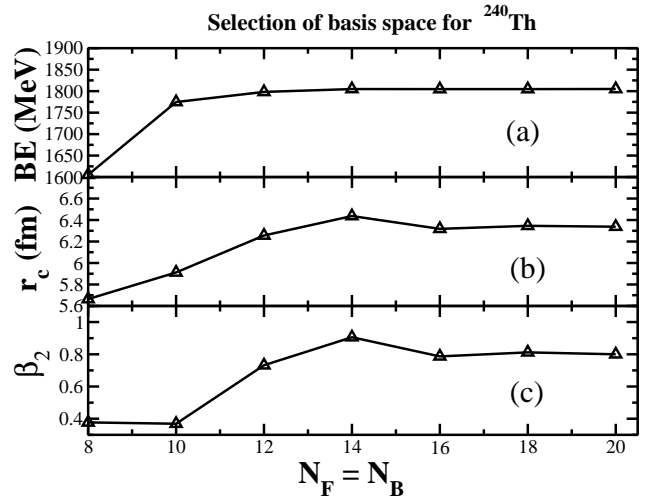}
\caption{ The variation of calculated binding energy (BE),
 charge radii ($r_{c}$) and quadrupole deformation parameter 
($\beta_{2}$) are given with Bosonic and Fermionic basis. 
}
\label{bea1}
\end{figure}

For heavy nuclei, a large number of basis space $N_F$ and $N_B$ are needed
to get a converged solution. To reduce the computational time without
compromising the convergence of the solution, we have to choose an optimal
number of model space for both Fermion and Boson fields.
To choose optimal values for $N_F$ and $N_B$, we select $^{240}$Th as a
test case and increase the basis quanta from 8 to 20 step by step. The
obtained results of binding energy, charge radii and quadrupole deformation
parameter are shown in Fig.~\ref{bea1}. From our calculations, we notice an increment 
of 200 MeV in binding energy while going from $N_F=N_B=$8 to 10. This increment
in energy decreases while going to higher oscillator basis. For example,
change in energy is $\sim 0.2$ MeV  with a change of $N_F=N_B$ from
14 to 20 and the increment in $r_c$ values are 0.12 fm  respectively.  
Keeping in mind the increase in convergence time for larger quanta as well as
the size of the nuclei considered, we have finalized to use $N_F=N_B=20$ 
in our calculations to get a suitable convergent results, which is the
current accuracy of the present RMF models.

\subsection{Binding energies, charge radii and quadrupole deformation 
parameters}\label{sec6}

To be sure about the predictivity of our model, first of all we calculate
the binding energies (BE), charge radii $r_c$ and quadrupole deformation
parameter $\beta_2$ for some of the known cases. We have compared our results 
with the experimental data wherever available or with the Finite Range Droplet
Model (FRDM) of M\"oller et al. \cite{NDCC,moller97,moller95,angel13}. 
The results are displayed in Tables~\ref{tab1} and~\ref{tab2}.  From the tables, 
it is obvious that the calculated binding energies are  comparable 
with the FRDM as well as experimental values.  A further inspection of 
the tables reveal that the FRDM results are more closer to the data. 
This may be due to the fitting of the FRDM parameters for almost all 
known data. However, in case of most RMF parametrizations, the
constants are determined by using few spherical nuclei data along with 
certain nuclear matter properties. Thus the prediction of the RMF 
results are considered to be reasonable, but not excellent.

Ren et al.~\cite{ren01,ren03} have reported that the ground state of several
superheavy nuclei are highly deformed states. 
Since, these are 
very heavy isotopes, the general assumption is that the ground state most
 probably remains in deformed configuration (liquid drop picture).
When these nuclei excited either by a thermal neutron or by any 
other means, it's intrinsic excited state becomes extra-ordinarily deformed 
and attains the scission point before it goes to fission. This can also be 
easily realized from the potential energy surface (PES) curve. Our calculations
agree with the prediction of Ren et al. for other superheavy region of the mass 
table.  However, this conclusion is contradicted by~\cite{soba91}. According 
to him, the ground state of superheavy nuclei either spherical or normally 
deformed. 

\begin{table*}
\hspace{0.1 cm}
\caption{The calculated binding energies BE, quadrupole deformation parameter
$\beta_{2}$, rms radii for the ground states and few selective intrinsic 
excited state of U isotopes, using  RMF formalism with NL3 parameter set. The experimental and FRDM data~\cite{NDCC,moller97,moller95,angel13} are also 
included in the table. See the text for more details.}
\renewcommand{\tabcolsep}{0.1 cm}
\renewcommand{\arraystretch}{1.0}
{\begin{tabular}{|c|c|c|c|c|c|c|c|c|c|c|c|c|c|c|c|c|c|c|c|c|c|c|}
\hline
& \multicolumn{6}{ |c| }{RMF (NL3)} &\multicolumn{2}{ |c| }{FRDM}& \multicolumn{3}{ |c| }{Experiment}  \\
\cline{2-6}\cline{6-7}\cline{7-10}
\hline
Nucleus &       $r_n$   &       $r_p$   &       $r_{rms}$       &       $r_{ch}$        &       $\beta_2$       &        BE (MeV)       &        BE(MeV)        &$\beta_2$&     $r_{ch}$        &       $\beta_2$       &        BE (MeV)       \\
\hline
&&&&&&&&&&&\\
$^{216}$U       &       5.762   &       5.616   &       5.700     &       5.673   &       0       &       1660.5  &       1649.0  &-0.052&                &               &               \\
        &       6.054   &       5.946   &       6.008   &       5.999   &       0.608     &       1650.8  &&              &               &               &               \\
        &               &               &               &               &               &               &       &       &               &               &               \\
$^{218}$U       &       5.789   &       5.625   &       5.721   &       5.682   &       0       &       1678.0  &       1666.7  &0.008&         &               &       1665.6  \\
        &       6.081   &       5.957   &       6.029   &       6.011   &       0.606     &       1666.9  &       &       &               &               &               \\
        &               &               &               &               &               &               &       &       &               &               &               \\
$^{220}$U       &       5.819   &       5.641   &       5.745   &       5.698   &       0       &       1692.2  &       1681.2  &0.008&         &               &       1680.8  \\
        &       6.109   &       5.971   &       6.052   &       6.025   &       0.605     &       1682.6  &&              &               &               &               \\
        &               &               &               &               &               &               &&              &               &               &               \\
$^{222}$U       &       5.849   &       5.661   &       5.772   &       5.717   &       0       &       1705.1  &       1695.7  &0.048&         &               &       1695.6  \\
        &       6.142   &       5.990    &       6.079   &       6.043   &       0.611    &       1697.9  &&              &               &               &               \\
        &               &               &               &               &               &               &&              &               &               &               \\

$^{224}$U       &       5.878   &       5.681   &       5.798   &       5.737   &       0       &       1717.9  &       1710.8  &0.146&         &               &       1710.3  \\
        &       6.198   &       6.032   &       6.131   &       6.085   &       0.645    &       1712.8  &&              &               &               &               \\
        &               &               &               &               &               &               &&              &               &               &               \\
$^{226}$U       &       5.907   &       5.701   &       5.824   &       5.757   &       0       &       1730.8  &       1724.7  &0.172&         &               &       1724.8  \\
        &       6.232   &       6.053   &       6.160    &       6.106   &       0.652    &       1727.4  &&              &               &               &               \\
        &               &               &               &               &               &               &&              &               &               &               \\
        &       5.935   &       5.721   &       5.850    &       5.776   &       0       &       1743.6  &&              &               &               &               \\
$^{228}$U       &       5.966   &       5.743   &       5.877   &       5.798   &       0.210    &       1741.7  &       1739.0  &0.191&         &               &       1739    \\
        &       6.259   &       6.068   &       6.182   &       6.120    &       0.651    &       1741.3  &&              &               &               &               \\
        &               &               &               &               &               &               &&              &               &               &               \\
        &          &          &          &          &          &         &&              &               &               &               \\
$^{230}$U       &       5.964   &       5.739   &       5.875   &       5.795   &       0       &       1756.0  &       1752.6  &0.199&         &       0.260    &       1752.8  \\
        &       6.000       &       5.765   &       5.907   &       5.821   &       0.234    &       1755.4  &&              &               &               &               \\
        &       6.293   &       6.091   &       6.213   &       6.143   &       0.658    &       1753.7  &&              &               &               &               \\
        &               &               &               &               &               &               &&              &               &               &               \\
        &          &          &          &          &          &         &&              &               &               &               \\
$^{232}$U       &       5.994   &       5.755   &       5.900     &       5.810    &       0       &       1766.8  &       1765.7  &0.207&         &       0.267    &       1765.9  \\
        &       6.033   &       5.785   &       5.935   &       5.840    &       0.251    &       1768.2  &&              &               &               &               \\
        &       6.364   &       6.167   &       6.286   &       6.218   &       0.712    &       1766.8  &&              &               &               &               \\
        &               &               &               &               &               &               &&              &               &               &               \\
        &          &          &          &           &          &         &&              &               &               &               \\
$^{234}$U       &       6.021   &       5.767   &       5.923   &       5.823   &       0       &       1776.4  &       1778.2  &0.215& 5.829   &       0.265    &       1778.6  \\
        &       6.065   &       5.803   &       5.963   &       5.858   &       0.267    &       1780.3  &&              &               &               &               \\
        &       6.415    &       6.209   &       6.334   &       6.260   &       0.738    &       1778.2  &&              &               &               &               \\
        &               &               &               &               &               &               &&              &               &               &               \\
        &          &          &          &          &          &         &&              &               &               &               \\
$^{236}$U       &       6.092   &       5.819   &       5.987   &       5.874   &       0.276    &       1791.7  &       1790.0  &0.215& 5.843   &       0.272    &       1790.4  \\
        &       6.446   &       6.230   &       6.363   &       6.281   &       0.744    &       1789.4  &&              &               &               &               \\
        &               &               &               &               &               &               &&              &               &               &               \\
        &          &          &          &          &          &         &&              &               &               &               \\
$^{238}$U       &       6.124   &       5.838   &       6.015   &       5.892   &       0.283    &       1802.5  &       1801.2  &0.215& 5.857   &       0.272    &       1801.7  \\
        &       6.488   &       6.263   &       6.402   &       6.314   &       0.763    &       1800.4  &&              &               &               &               \\
        &               &               &               &               &               &               &&              &               &               &               \\
\hline
%\hline
\end{tabular}\label{tab1} }
\end{table*}

%\end{document}

\begin{table*}
\hspace{0.1 cm}
\caption{Same as Table I,  but for Th isotopes.}
\renewcommand{\tabcolsep}{0.1 cm}
\renewcommand{\arraystretch}{1.0}
{\begin{tabular}{|c|c|c|c|c|c|c|c|c|c|c|c|c|c|c|c|c|c|c|c|c|c|}
\hline
& \multicolumn{6}{ |c| }{RMF (NL3)} &\multicolumn{2}{ |c| }{FRDM}& \multicolumn{3}{ |c| }{Experiment}  \\
\cline{2-6}\cline{6-7}\cline{7-10}
\hline
Nucleus &       $r_n$   &       $r_p$   &       $r_{rms}$       &       $r_{ch}$        &       $\beta_2$       &       BE (MeV)        &       BE (MeV)        &$\beta_2$&     $r_{ch}$        &       $\beta_2$       &       BE(MeV) \\
\hline
&&&&&&&&&&&\\
$^{216}$Th      &       5.781   &       5.594   &       5.704   &       5.651   &       0       &       1673.5  &       1663.6  &0.008&         &               &       1662.7  \\
        &       6.034    &       5.897   &       5.977   &       5.951   &       0.567    &       1663.8  &&              &               &               &               \\
        &               &               &               &               &               &               &&              &               &               &               \\
$^{218}$Th      &       5.812   &       5.611   &       5.730    &       5.667   &       0       &       1686.5  &       1677.2  &0.008&         &               &       1676.7  \\
        &       6.105   &       5.959   &       6.045   &       6.013   &       0.616    &       1678.2  &&              &               &               &               \\
        &               &               &               &               &               &               &&              &               &               &               \\
$^{220}$Th      &       5.842   &       5.631   &       5.757   &       5.687   &       0       &       1698.1  &       1690.2  &0.030&         &               &       1690.6  \\
        &       6.140   &       5.983   &       6.076   &       6.036   &       0.624    &       1692.8  &&              &               &               &               \\
        &               &               &               &               &               &               &&              &               &               &               \\
$^{222}$Th      &       5.873   &       5.651   &       5.784   &       5.707   &       0       &       1709.7  &       1704.6  &0.111&         &       0.151   &       1704.2  \\
        &       6.174   &       6.007   &       6.107   &       6.060   &       0.631    &       1706.1  &&              &               &               &               \\
        &               &               &               &               &               &               &&              &               &               &               \\

$^{224}$Th      &       5.902   &       5.672   &       5.81    &       5.728   &       0       &       1721.4  &       1717.4  &0.164&         &       0.173   &       1717.6  \\
        &       6.222   &       6.021   &       6.142   &       6.074   &       0.640    &       1718.9  &&              &               &               &               \\
        &               &               &               &               &               &               &&              &               &               &               \\
        &          &          &          &          &         &        &         &&         &         &        \\
$^{226}$Th      &       5.931   &       5.692   &       5.837   &       5.748   &       0       &       1733.0  &1729.9& 0.173             &             &    0.225           &       1730.5        \\
        &       6.25    &       6.036   &       6.166   &       6.089   &       0.642    &       1731.9  &&              &               &               &               \\
        &               &               &               &               &               &               &&              &               &               &               \\
$^{228}$Th         &       5.955    &       5.710   &       5.859   &       5.766    &       0  &       1743.9  &1742.5&0.182              &  5.748             &    0.229            &  1743.0             \\
     &       5.989   &       5.729   &       5.888   &       5.785   &       0.227       &       1744.5  &         &&    &         &         \\
        &       6.292   &       6.065   &       6.203   &       6.118   &       0.661    &       1743.4  &&              &               &               &               \\
        &               &               &               &               &               &               &&              &               &               &               \\
        &       &          &          &          &          &         &&              &               &               &               \\
$^{230}$Th      &       5.990    &       5.727   &       5.888   &       5.783   &       0       &       1754.2  &       1754.6  &0.198& 5.767   &       0.246   &       1755.1  \\
        &       6.026   &       5.751   &       5.920    &       5.807   &       0.232    &       1756.0  &&              &               &               &               \\
        &       6.315   &       6.111    &       6.236   &       6.163   &       0.671    &       1753.1  &&              &               &               &               \\
        &               &               &               &               &               &               &&              &               &               &               \\
        &          &         &          &          &         &         &&              &               &               &               \\
$^{232}$Th      & 6.060      &  5.773       & 5.950      & 5.828   &0.251        &1767.0        &1766.2& 0.207             & 5.784              &   0.248             &  1766.7             \\
        &   6.240        &     6.010     &    6.151      &   6.063       &  0.681         & 1765.0        &         &&    &         &         \\
        &          &          &          &          &           &         &&              &               &               &               \\
        &               &               &               &               &               &               &&              &               &               &               \\
%        &           &          &          &           &          &         &&              &               &               &               \\
%      &      &      &     &      &      &     &&              &               &               &               \\
$^{234}$Th        &       6.093   &       5.793   &       5.979   &       5.848   &       0.269    &       1777.5  &       1777.2  &0.215&         &       0.238   &       1777.6  \\
        &          &          &          &          &           &         &&              &               &               &               \\
%        &               &               &               &               &               &               &&              &               &               &               \\
%        &          &          &           &          &          &         &&              &               &               &               \\
$^{236}$Th      &6.122       &5.812     &6.006     &5.866      & 0.272   & 1787.6    &       1787.6  &0.215&         &               &       1788.1  \\
%        &          &          &           &          &           &          &&              &               &               &               \\
%        &          &          &         &          &           &        &&              &               &               &               \\
%        &               &               &               &               &               &               &&              &               &               &               \\
        &          &         &          &          &           &         &&              &               &               &               \\
$^{238}$Th      &6.152       & 5.832      &6.033    & 5.887     & 0.281      & 1797.5  &       1797.7  &0.224&         &               &       1797.8  \\
        &           &           &          &          &           &          &&              &               &               &               \\
%        &         &         &          &          &           &        &&              &               &               &               \\
%        &               &               &               &               &               &               &&              &               &               &               \\
%        &          &          &          &          &           &         &         &  &       &               &               \\
$^{240}$Th      & 6.180        &5.846     & 6.057   &5.901        &0.292         & 1806.6       &1807.2& 0.224             &               &               &               \\
        &           &          &          &         &          &        &&              &               &               &               \\
%        &           &          &          &          &            &         &&              &               &               &               \\
%        &               &               &               &               &               &               &&              &               &               &               \\
\hline
%\hline
\end{tabular}\label{tab2} }
\end{table*}
In some cases of U and Th isotopes, we get more than one solution. The solution 
corresponding to the maximum binding energy is the ground state configuration and
all other solutions are the intrinsic excited states. In some cases, the ground
state binding energy does not match with the experimental data. However, the
binding energy, whose quadrupole deformation parameter $\beta_2$ is closer to the 
experimental data or to the FRDM value matches well with each other. For example,
binding energies of $^{236}$U are 1791.7, 1790.0 and 1790.4 MeV with RMF, FRDM 
and experimental data, respectively and the corresponding $\beta_2$ are 0.276, 
0.215 and 0.272.  
Similar to the binding energy, we get comparable $\beta_2$ and charge radius $r_c$
of RMF results with the FRDM and experimental values.
% In some cases, we have also 
%noticed oblate and super-deformed states, which are slightly above the ground state configuration.   

\subsection{Potential energy surface (PES)}\label{sec7}

In late 1960's, the structure of potential energy surface (PES) has 
been renewed interest for it's role in nuclear fission process. 
In majority of PES for actinide nuclei, there exists a 
second maximum, which split the fission barrier into inner and outer 
segments ~\cite{lynn80}.
It has also a crucial role for the characterization of ground state, intrinsic
excited state, occurrence of the shape coexistence, radioactivity,
spontaneous and induced fission. 
 The structure of the potential energy 
surface is defined mainly from the shell structure which is strongly related
to the distance between the mass centers of the nascent fragments.
The macroscopic-microscopic liquid drop theory has been given a key 
concept of fission, where the surface energy is the form of collective 
deformation of the nucleus. 

\begin{figure}[ht]
\includegraphics*[width=1.1\columnwidth]{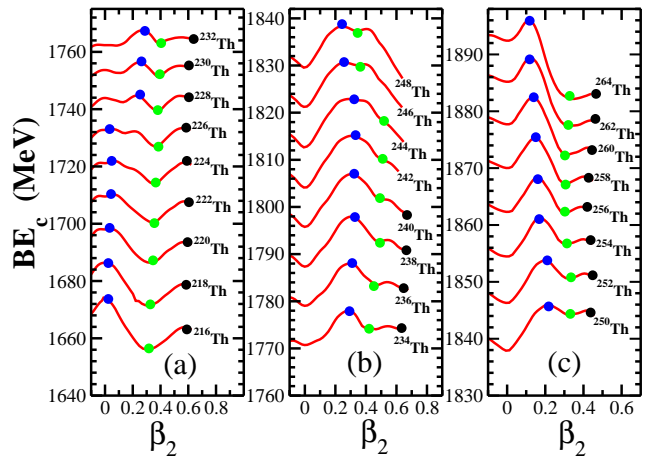}
\caption{(Color online) The potential energy surface is a function of 
quadrupole deformation parameter ($\beta_{2}$) for Th isotopes. The difference
between blue and green dots represents first fission barrier heights B$_{f}$
(in MeV). See text for details.}
\label{bea2}
\end{figure}

\begin{figure}[ht]
\includegraphics*[width=1.1\columnwidth]{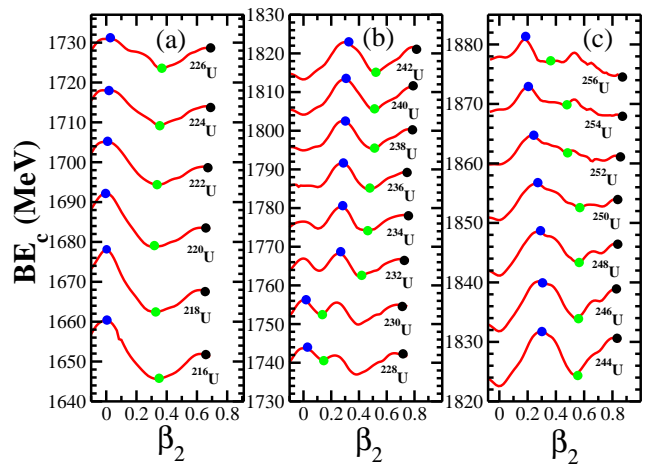}
\caption{(Color online) Same as Fig. 2, but for U isotopes.}
\label{bea3}
\end{figure}

In Figs.~\ref{bea2} and ~\ref{bea3} we have plotted the PES for some selected isotopes of
Th and U nuclei. The constraint binding energy BE$_{c}$ versus the quadrupole 
deformation parameter $\beta_2$ are shown. 
A nucleus undergoes fission process, when the nucleus becomes highly elongated
along an axis.  This can be done in a simplest way by modifying the 
single-particle potential with the help of a constraint, i.e., the 
Lagrangian multiplier $\lambda$.
Then, the system becomes more or less compressed depending
on the Lagrangian multiplier $\lambda$. 
In other word, in a constraint calculation, we minimize 
the expectation value of the Hamiltonian $<H'>$ instead of $<H>$ which 
are related to each other by the following 
relation~\cite{patra09,flocard73,koepf88,fink89,hirata93}:
\begin{eqnarray}
H^{'}=H-\lambda Q,\qquad {with} \qquad Q=r^{2}Y_{20}(\theta, \phi),
\end{eqnarray}
where, $\lambda$ is fixed by the condition $<Q>_\lambda$ = $Q_{0}$.

\begin{table}
\hspace{0.1 cm}
\caption{First fission barrier heights B$_{f}$ (in MeV) of some
even-even actinide nuclei from RMF(NL3) calculations compared
with FRDM and experimental data~\cite{moller95}.}
\renewcommand{\tabcolsep}{0.2 cm}
\renewcommand{\arraystretch}{1.2}
{\begin{tabular}{|c|c|c|c|c|}
\hline
 \multicolumn{1}{ |c| }{Nucleus} &\multicolumn{1}{ |c| }{B$^{cal.}_{f}$ }& 
\multicolumn{1}{ |c| }{B$^{FRDM}_{f}$~\cite{moller95}}&\multicolumn{1}{ |c| }{B$^{exp.}_{f}$~\cite{moller95}} \\ 
\cline{1-3}
\hline
$^{228}$Th & 5.69 &  7.43  &6.50\\
$^{230}$Th & 5.25 &  7.57  &7.0\\
$^{232}$Th & 4.85 &  7.63  &6.30\\
$^{234}$Th & 4.34 &  7.44  &6.65\\
$^{232}$U & 5.65 &  6.61  &5.40\\
$^{234}$U & 6.30 &  6.79  &5.80\\
$^{236}$U & 6.64 &  6.65  &5.75\\
$^{238}$U & 7.15 &  4.89  &5.90\\
$^{240}$U & 7.66 &  5.59  &5.80\\
\hline
\end{tabular}\label{tab3}}
\end{table}

Usually, in an axially deformed constraint calculation for a nucleus,
we see two maxima in the PES diagram, (i) prolate and (ii) oblate or
spherical. However, in some cases, more than two maxima are also seen.
 If the ground state energy is distinctly more than other maxima, then
the nucleus has a well defined ground state configuration. On the other
hand, if the difference in binding energy between two or three maxima is 
negligible, then the nucleus is in shape co-existence configuration. In such a
case, a configuration mixing calculation is needed to determine the
ground state solution of the nucleus, which is beyond the scope of the
present calculation. It is to be noted here that in a constraint calculation,
the maximum binding energy (major peak in the PES diagram) corresponds to
the ground state configuration and all other solutions (minor peaks in the
PES curve) are the intrinsic excited states. 

The fission barrier $B_f$ is an important quantity to study the properties of 
fission reaction.  We calculate the fission barrier from the PES 
curve for some  selected even-even nuclei, which are displayed in 
Table~\ref{tab3}. From the table, it can be
seen that the fission barrier for $^{228}$Th comes out to be 5.69 MeV 
comparable to the FRDM and experimental values of $B_f$ = 7.43  and 6.50 MeV,
respectively. Similarly, the calculated $B_f$ of $^{232}$U is 
5.65 MeV, which also agree  well with the experimental data 5.40 MeV. In 
some cases, the fission barrier height is 1$-$2 MeV lower or higher than 
the experimental data. 
The double-humped fission barrier in all these cases are reproduced.
Similar type of calculations are also done in Refs.~\cite{meng06,bing12,nan14,zhao15}. 

In nuclei like $^{228-230}$Th and $^{228-234}$U, we find three maxima. Among 
these maxima, two of them are found at normal deformation 
(spherical and normal prolate), but the third one is situated far away, i.e., at
relatively large quadrupole deformation. With a careful inspection, one can 
also see that one of them (mostly the peak nearer to the spherical region) 
is not strongly pronounced and can be ignored in certain cases. This third 
maximum separate 
the second barrier with a depth of  1 - 2 MeV, responsible for the
formation of resonance state, which are observed experimentally\cite{back72}.
Some of the uranium isotopes $^{216-230}$U, the ground states are predicted
to be spherical in RMF formalism agreeing with the FRDM results. The other
isotopes of the series $^{232-256}$U are found to be prolate ground state
matching with the experimental data.
Similarly, the thorium nuclei $^{216-226}$Th are spherical in shape and 
$^{228-264}$Th are prolate ground configuration.
In addition to these shapes, we also notice sallow regions in the 
PES curves of both Th and U isotopes. These fluctuation in the PES
curves could be due to the limitation of mean field approximation
and one needs a theory beyond mean field to
over come such fluctuations. For example, the Generator Coordinate Method
or Random Phase Approximation could be some improved formalism to take 
care of such effects~\cite{brink68}. Beyond the second hump, we find the
PES curve goes down and down, which never ups again. This is the process
of the liquid drop gets more and more elongation and reaches to the
fission stage. The PES curve, from which it starts downing is marked the
scission points which are shown by the black dot in some of the PES curves
of Figs.~\ref{bea2} and ~\ref{bea3}.

\subsection{Evolution of single-particle energy with deformation}\label{sec8}

\begin{figure}[ht]
\includegraphics*[width=1.1\columnwidth]{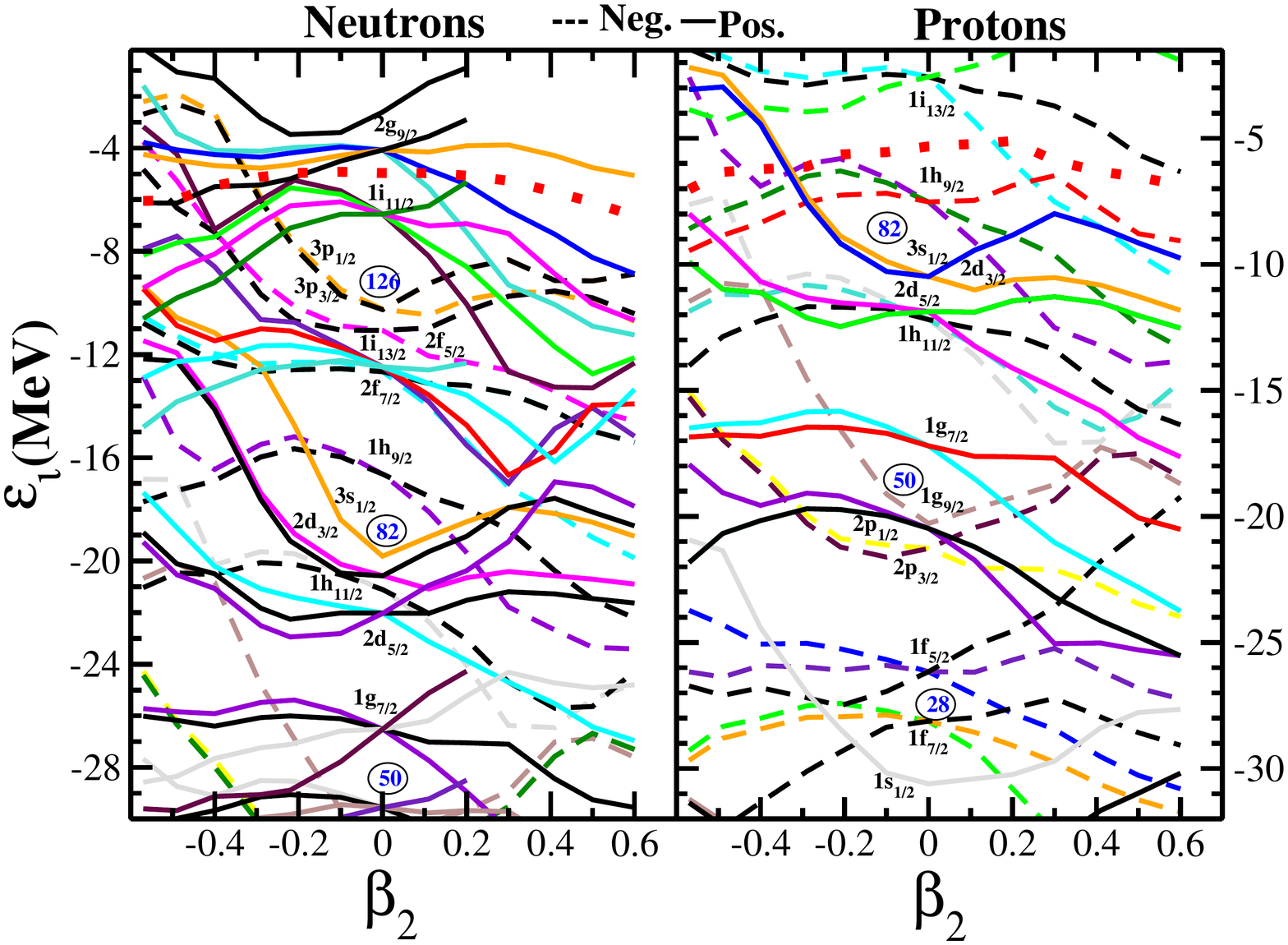}
\caption{(Color online) Single-particle energy levels for $^{232}$Th as a 
function of quadrupole deformation parameter $\beta_{2}$. The Fermi levels are 
denoted by thick dotted(red) curve.
}
\label{bea4}
\end{figure}

\begin{figure}[ht]
\includegraphics*[width=1.1\columnwidth]{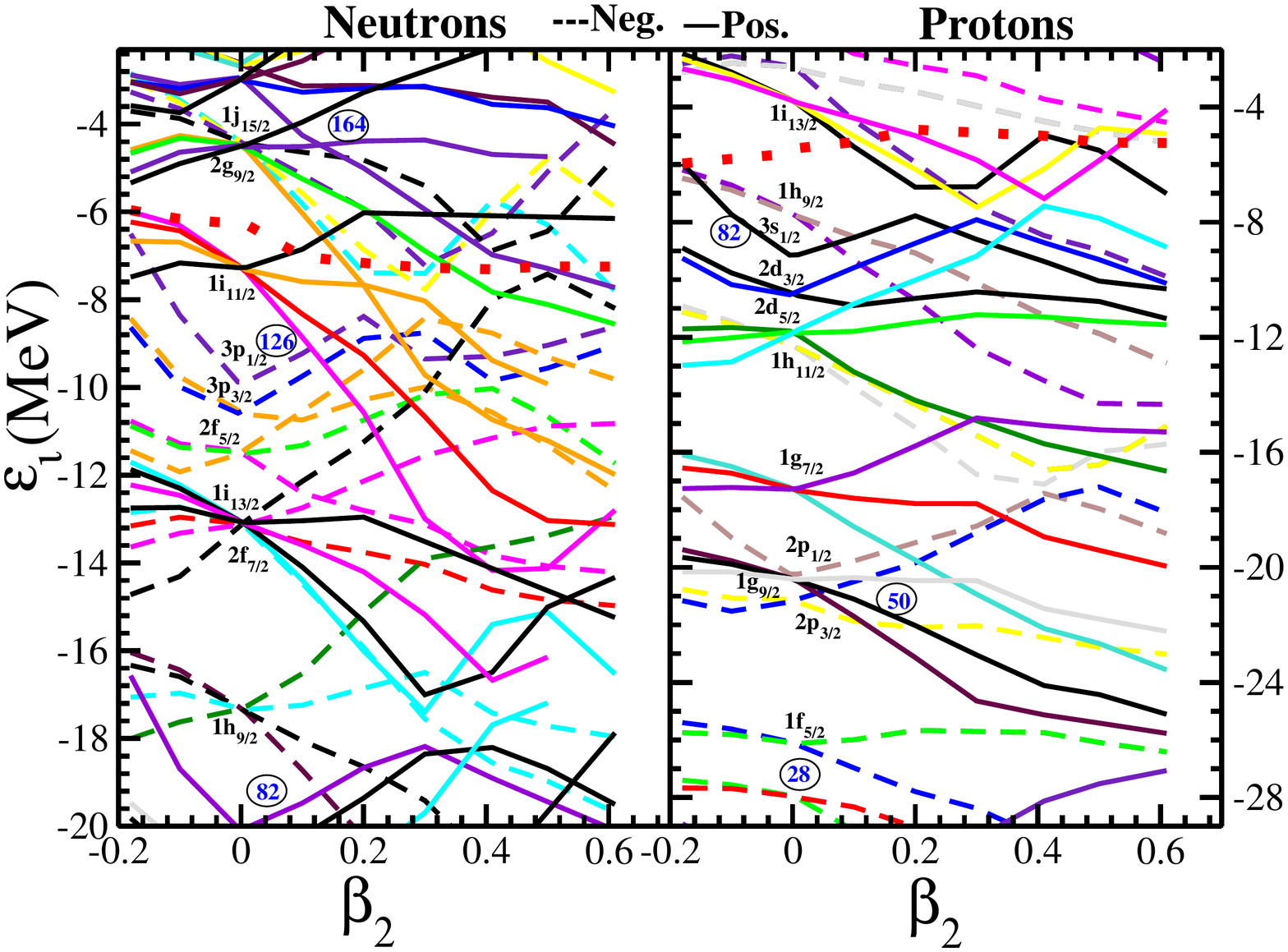}
\caption{(Color online) Same as Fig. 4, but for $^{236}$U nucleus.
}
\label{bea5}
\end{figure}

In this subsection, we evaluate the neutron and proton single-particle 
energy levels for some selected Nilssion orbits 
with different values of deformation parameter $\beta_{2}$ using the 
constraint calculations. The results are given in Figs.~\ref{bea4} and 
~\ref{bea5}, explain the
origin of the shape change along the $\alpha$-decay chains of the thorium
and uranium isotopes. The positive parity orbit is the solid
line, negative parity orbit is dash line and the dotted line (red colour) 
indicates the Fermi energy for $^{232}$Th and $^{236}$U.

For small Z nuclei, the electrostatic repulsion is very weak but at higher 
value of Z (superheavy nuclei), the electrostatic repulsion is much stronger
that the nuclear liquid drop becomes unstable to surface distortion
~\cite{mayer66} and fission. 
In such nucleus, the single-particle density is very large and the energy 
separation is small, which determines the shell stabilizes the unstable
Coulomb repulsion. This effect is clear for heavy elements approaching 
N=126 with the gap between 3p$_{1/2}$ and 1i$_{11/2}$ of about 2-3 MeV, in
the neutron single-particle of $^{236}$U and $^{232}$Th.
In both the figures, the neutron single particle energy level
1i$_{13/2}$ lies between 2f$_{7/2}$ and 2f$_{5/2}$ creating a distinct 
shell gap at N=114. 
In $^{232}$Th and $^{236}$U, with increasing deformation the opposite 
parity levels of 2g$_{9/2}$ and 1j$_{15/2}$ come closer to each other, 
which are far apart in the spherical solution. This gives rise to the
parity doublet phenomena \cite{singh14,kumar15,hab02}.

\section {Mode of decays}\label{sec9}

In this section, we will discuss about various mode of decays encounter
by superheavy nuclei both in the $\beta$-stability line as well as away
from it. This is important, because the utility of superheavy and mostly 
the  nuclei which are away from stability lines depend very much
on their life time. For example, we do not get $^{233}$U and $^{239}$Pu in
nature, because of their short life time, although these two nuclei are
extremely useful for energy production. That is why $^{235}$U is the most
necessary isotope in the uranium series for its thermally fissile nature
in the energy production in fission process both for civilian as well as
military use. The common mode of instability for such heavy nuclei are
spontaneous fission, $\alpha$-,  $\beta$- and $cluster$-decays. All these
decays depend on the neutron to proton ratio as well as the number of nucleons
present in the nucleus.

\subsection{$\alpha$- and $\beta$-decays half-lives}\label{sec10}

In the previous papers\cite{bk06,patra07}, we have analyzed the densities of 
 nuclei in a more detailed manner. From this analysis, we concluded that 
there is no visible cluster either in the ground or in the excited intrinsic 
states. The possible clusterizations are the $\alpha$-like matter at the 
interior and neutron-rich matter at the exterior  region of the normal 
and neutron-rich superheavy nuclei, respectively. Thus, the possible mode 
of decays are the
$\alpha$-decay for $\beta$-stable nuclei and $\beta^-$-decay for neutron-rich
isotopes. To estimate the stability of such nuclei, we have to calculate
the $\alpha$-decay $T_{1/2}(\alpha)$ and the $\beta$-decay $T_{1/2}(\beta)$ 
half-lives times.

\subsubsection { The Q$_\alpha$ energy and $\alpha$-decay half-life 
              T$_{1/2}^{\alpha}$ }\label{sec11}

To calculate the $\alpha$-decay half-life $T_{1/2}^{\alpha}$, one
has to know the $Q_{\alpha}$ energies of the nucleus. This can be estimated
by knowing the binding energies (BE) of the parents, daughter and the binding
energy of the $\alpha$-particle, i.e., the BE of $^{4}$He. The binding
energies are obtained from experimental data wherever available and from 
other mass formulae as well as relativistic mean field Lagrangian as we
have discussed earlier in this paper \cite{patra23}. The $Q_{\alpha}$
energy is evaluated by using the relation:
\begin{eqnarray}
Q_{\alpha}(N,Z)&=& BE(N, Z)- BE(N-2,Z-2)\nonumber\\
&-& BE(2,2)
\end{eqnarray}
Here, $BE(N, Z)$, $BE(N-2,Z-2)$ and $BE(2,2)$ are the binding energies
of the parent, daughter and $^4$He nuclei (BE= 28.296 MeV)
with neutron number N and proton number Z.
\begin{figure}[ht]
\includegraphics*[width=1.\columnwidth]{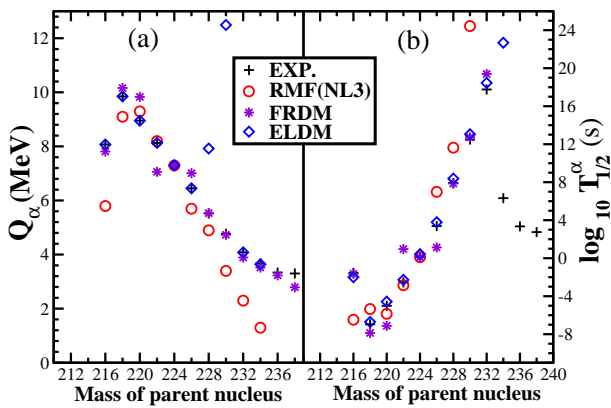}
\caption{(Color online) Q$_\alpha$  and half-life time
$T^{\alpha}_{1/2}$ of the $\alpha$-decay chain for
Th isotopes are calculated using RMF, FRMD~\cite{moller97,moller95}
, ELDM~\cite{sb02} and compared with the experiment~\cite{NDCC}.}
\label{bea11}
\end{figure}

\begin{figure}[ht]
\includegraphics*[width=1.\columnwidth]{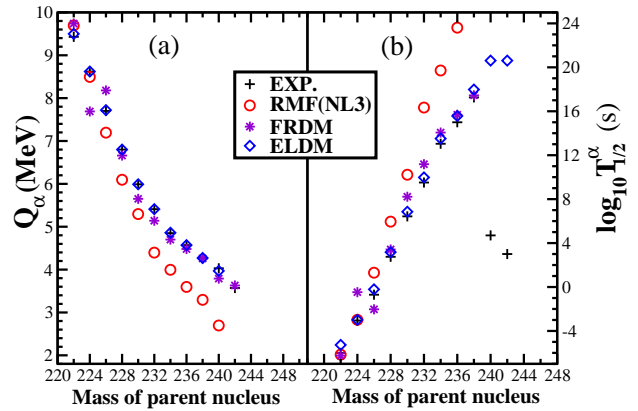}
\caption{(Color online) Same as Fig. 6, but for U.}
\label{bea12}
\end{figure}
Knowing the $Q_{\alpha}-$values of nuclei, we roughly estimate the 
$\alpha$-decay half-lives $log_{10}T_{1/2}^{\alpha} (s)$ of various nuclei 
using the phenomenological formula of Viola and Seaborg \cite{viol01}:
\begin{equation}
log_{10}T^{\alpha}_{1/2}(s)=\frac {(aZ-b)}{\sqrt{Q_{\alpha}}}-(cZ+d) + 
h_{log}.
\end{equation}
The value of the parameters $a$, $b$, $c$ and $d$ are taken from the
recent modified parametrizations of Sobiczewski et al.
\cite{sobi89}, which are $a$ = 1.66175; $b$ = 8.5166; $c$ = 0.20228; 
$d$ = 33.9069.
The quantity $h_{log}$ accounts for the hindrances associated
with the odd proton and neutron numbers as given by
Viola and Seaborg~\cite{viol01}, namely

$h_{log}=\begin{array}{ll}
           0, &  \mbox{ Z and N even}\\
        0.772,&  \mbox{ Z odd and N even}\\ 
        1.066,&  \mbox{ Z even and N odd}\\ 
        1.114,&  \mbox{ Z and N odd}. 
\end{array}$

The $Q_{\alpha}-$values obtained from RMF calculations for Th and U isotopes 
are shown in Figs.~\ref{bea11} and~\ref{bea12}. Our results also compared 
with other theoretical predictions \cite{moller97,sb02} and experimental 
data \cite{angel13}. The agreement of RMF results with others as well as 
with experiment is pretty well.
Although, the agreement in $Q_{\alpha}-$value is quite good, one has to note
that the $T^{\alpha}_{1/2}(s)$ values may vary a lot, because of the exponential
factor in it. That is why it is better to compare  
$log_{10}T^{\alpha}_{1/2}(s)$ instead of  $T^{\alpha}_{1/2}(s)$. These values 
are compared in the right panel of Figs.~\ref{bea11} and~\ref{bea12}.
 We notice, our prediction
matches well with other calculations as well as experimental data.

Further, a careful analysis of $log_{10}T^{\alpha}_{1/2}$ (in seconds) for 
even-even thorium, the $Q_{\alpha}-$value decreases 
with increase of mass number A of parent nucleus. The $Q_{\alpha}$ energy 
of Th isotopes given by Duarte et al. \cite{sb02} deviates a lot, when mass of 
the parent nucleus reaches to A=230. The corresponding 
$log_{10}T^{\alpha}_{1/2}$ increases almost monotonically linearly with an
increase of mass number of the same nucleus. The experimental values of 
$log_{10}T^{\alpha}_{1/2}$ deviate a lot in the heavy mass region, 
(with parent nuclei 234-238). Similar situation is found
in case of uranium isotopes also which are shown in Fig.~\ref{bea12}.

\subsection{$\beta$-decay}\label{sec12}

As we have discussed, the prominent mode of instability of neutron-rich
Th and U nuclei is the $\beta$-decay, and we have given an estimation of such
decay in this subsection. Actually, the $\beta$-decay life time should be
evaluated in a microscopic level, but in this paper, it is beyond the scope.
Here we have used the empirical formula of Fiset and Nix \cite{Fiset72}, which
is defined as:
\begin{figure}[ht]
\begin{center}
\includegraphics*[width=1.\columnwidth]{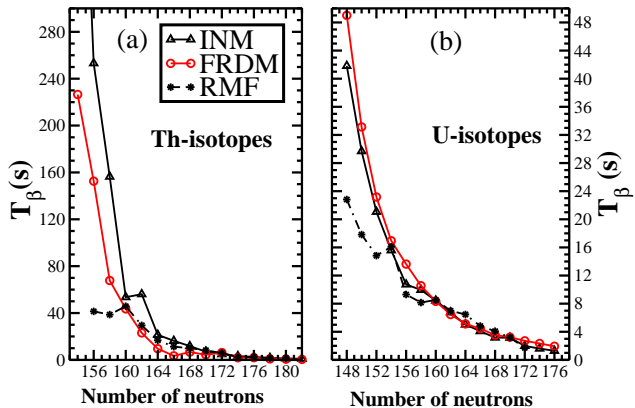}
\caption{(Color online) The $\beta$-decay half life for Th and U isotopes are
calculated using the formula of Fiset and Nix~\cite{Fiset72} [eq. (24)]. The 
ground state binding energies are 
taken from FRDM~\cite{moller97}, INM~\cite{nayak99} and RMF models.
}
\label{bea15}
\end{center}
\end{figure}
\begin{eqnarray}
T{_\beta} = (540 \times 10^{5.0})\frac {m{_e}^{5}} {\rho_{d.s.} (W{_\beta}^{6}-m{_e}^{6})}   s.  
\end{eqnarray}
Similar to the $\alpha$-decay, we evaluate the $Q_{\beta}$-value for
Th and U series using the relation $Q{_\beta} = BE(Z+1,A) - B(Z,A)$
and $W{_\beta} = Q{_\beta} + m{_e}^{2}$. Here, $\rho_{d.s.}$ is the 
average density of states in the daughter nucleus
(e$^{-A/290}$ $\times$ number of states within 1 MeV of ground state).
To evaluate the bulk properties, such as binding energy of odd-Z nuclei, 
we used the Pauli blocking prescription as discussed in Section~\ref{sec2}.
The obtained results are displayed in Fig.~\ref{bea15} for both Th and U 
isotopes.
From the figure, it is clear that for neutron-rich Th and U nuclei, the
prominent mode of decay is $\beta$-decay. This means, once the neutron-rich
thermally fissile isotope is formed by some artificial mean in laboratory
or naturally in supernovae explosion, immediately it undergoes $\beta$-decay.
In our rough estimation, the life time of $^{254}$Th and $^{256}$U, which
are the nuclei of interest has tens of seconds. If this prediction of
time period is acceptable, then in nuclear physics scale, is reasonably a
good time for further use of the nuclei. It is worthy to mention here that
thermally fissile isotopes of Th and U series are with neutron number
N=154-172 keeping N=164 in the middle of the island. So, in case of the
short life time of $^{254}$Th and $^{256}$U, one can choose a lighter isotope
of the series for practical utility.  

\subsection{Limitations of the model}\label{sec13}

Before drawing the concluding remarks, it is important to
mention few points about the limitations of our present approach. When we 
compare our calculated results with the experimental data, although we get
satisfactory results, some time we do not get excellent agreement and the
main possible reasons for the discrepancy of RMF with experimental values 
are given as:\\

(1) In RMF formalism we are working in the mean field approximation of the 
meson field. In this approximation, we are neglecting the vacuum 
fluctuation, which is an indispensable part of the relativistic formalism. 
In calculating the nucleonic dynamics, we are neglecting the negative energy 
solution that means, we are working in the no sea approximation~\cite{rufa86}. 
It is already discussed that the no-sea approximation and quantum 
fluctuation can improve the results upto a maximum of 20\% ~\cite{zhu91} for
very light-nuclei. Therefore, the mean field is not a good approach for the 
light region of the periodic table. However, for the heavy masses, this mean
field approach is quite good and can be used for any practical purpose. \\

(2) In order to solve the nuclear many body system, here we used the Hartee 
formalism and neglect Fock term, which corresponds to the exchange correlation.\\ 

(3) To take care of the pairing correlation, we have used BCS type pairing
approach. This gives  good results for the nuclei near the $\beta$-stability 
line, but it fails to incorporate properly the pairing correlation for 
the nuclei away from the 
$\beta$-stability line and superheavy nuclei~\cite{karatzikos10}. Thus a 
better approach like Hartree-Fock-Bogoliubov~\cite{ring80,ring96} type pairing 
co-relation is more suitable for the present region. \\

(4) Parametrization  plays an important role in improvising the results. 
The constants in RMF parametrizations, are determined by fixing the 
experimental data for few spherical nuclei. We expect that the results may
be improved by re-fitting the force parameters for more number of nuclei,
including the deformed isotopes.\\

(5) The basic assumption in the RMF theory is that 
two nucleons interact  with each other through the exchange of 
various mesons. There is no direct inclusion of 3-body or higher body effects.
This effect is taken care partially by including the self-coupling of mesons 
and in recent relativistic approach various cross-couplings are added because of
their importance.\\

(6) Although, there are various mesons are observed experimentally, few of them 
are taken into account in the nucleon-nucleon interaction. Contribution of 
some of them are prohibited due to symmetry reason and many are neglected 
due to their negligible contributions, because of heavy mass. However, some 
of them has substantial contribution to the properties of nuclei, specially 
when the neutron-proton asymmetry is more, such as $\delta$-meson~\cite{kubis97,sing14}.
\\

(7) It is to be noted that the origin of $\alpha$-decay or $cluster$-decay 
phenomena are purely quantum mechanical process. Thus the quantum tunneling
plays an important role in such decay processes. The deviation of experimental
$\alpha$-decay life time from the calculated results obtained by the empirical
formula may not be suitable for such heavy nuclei, which are away from
the stability line and more involved quantum mechanical treatment is needed 
for such cases.

\section{Conclusions}\label{sec14}
In summary, we did a thorough structural study of the recently predicted 
thermally fissile isotopes of Th and U series in the framework of 
relativistic mean field theory.  Although there are certain limitations 
of the present approach, the qualitative results will remain unchanged even
if the draw-back of the model taken into account. The heavier isotopes of 
these two nuclei
bear various shapes including very large prolate deformation at high 
excited configurations. The change in single-particle orbits along
the line of quadrupole deformation are analyzed and found parity doublet
states in some cases. Using an empirical estimation, we find that the 
neutron-rich isotopes of these thermally fissile nuclei are predicted to 
be stable against $\alpha$- and $cluster$-decays.
The spontaneous fission also does not occur, because the presence of large 
number of neutrons makes the fission barrier broader.  However, these 
nuclei are highly $\beta$-unstable. Our calculation predicts that the
$\beta$-life time is about tens of seconds for $^{254}$Th and $^{256}$U
and this time increases for nuclei with less neutron number, but thermally 
fissile. This finite life time of these thermally fissile isotopes could be
very useful for energy production in nuclear reactor technology. If these
neutron-rich nuclei use as nuclear fuel, the reactor will achieve critical
condition much faster than the normal nuclear fuel, because of the release
of large number of neutrons during the fission process.

\end{document}